\documentclass[english,aps,preprint]{revtex4}
\usepackage[T1]{fontenc}
\usepackage[latin9]{inputenc}
\usepackage{amssymb}
\usepackage{graphicx}

\makeatletter

\providecommand{\tabularnewline}{\\}

\@ifundefined{textcolor}{}
{%
 \definecolor{BLACK}{gray}{0}
 \definecolor{WHITE}{gray}{1}
 \definecolor{RED}{rgb}{1,0,0}
 \definecolor{GREEN}{rgb}{0,1,0}
 \definecolor{BLUE}{rgb}{0,0,1}
 \definecolor{CYAN}{cmyk}{1,0,0,0}
 \definecolor{MAGENTA}{cmyk}{0,1,0,0}
 \definecolor{YELLOW}{cmyk}{0,0,1,0}
}

\usepackage{babel}

\usepackage{babel}

\makeatother

\usepackage{babel}
\begin{document}

\title{Associated Production of Single Top Quark and W-boson Through Anomalous
Couplings at LHeC based $\gamma p$ Colliders}

\author{I. T. Cakir}

\email{ilkay.turkcakir@gmail.com}

\selectlanguage{english}%

\affiliation{Ankara University, Department of Physics, 06100, Ankara, Turkey }

\author{A. Senol}

\email{asenol@kastamonu.edu.tr}

\selectlanguage{english}%

\affiliation{Kastamonu University, Department of Physics, 37100, Kuzeykent, Kastamonu,
Turkey}

\affiliation{Abant Izzet Baysal University, Department of Physics, 14280, Bolu,
Turkey }

\author{A. T. Tasci}

\email{atasci@kastamonu.edu.tr}

\selectlanguage{english}%

\affiliation{Kastamonu University, Department of Physics, 37100, Kuzeykent, Kastamonu,
Turkey}
\begin{abstract}
We consider the production of a single top quark in association with
a $W$ boson at LHeC based $\gamma p$ collider. We compute the cross
section for the process $\gamma p\rightarrow WtX$ with the anomalous
$Wtb$ and $Wtb\gamma$ couplings. We find that the sensitivities
to anomalous couplings of top quark are shown to be comparable, even
better than the ones obtained from direct searches at hadron colliders.
\end{abstract}


\maketitle

\section{introduction}

The considerable successes of top quark physics have entered into
the field of precision measurements with the operation of Large Hadron
Collider (LHC). Due to the large mass, close to the scale of electroweak
symmetry breaking, the top quark is expected to be the most sensitive
to new physics beyond the Standard Model (SM). In the SM, the top
quark ($t$)-bottom quark ($b$)-$W$ boson interaction vertex is
defined by the strength $gF_{1L}/\sqrt{2}$ in which the coupling
$F_{1L}$ reduces to the quark mixing element $V_{tb}\simeq1$ at
tree level. Corrections to this coupling, as well as non-zero anomalous
couplings $F_{1R}$, $F_{2L}$ and $F_{2R}$ can be generated by the
new physics.

The direct constraints on the anomalous $Wtb$ couplings, using the
cross section measurement provided by CMS \cite{1} and ATLAS \cite{2}
for t- channel single top quark production and measurement of decay
asymmetries of top quark by ATLAS \cite{3}, are given in the region
(-0.55, 0.65) for $F_{1R},$ (0.55, 1.55) for $F_{1L},$ (-0.70, 0.25)
for $F_{2R}$ and (-0.60, 0.55) for $F_{2L}$ \cite{4}. The Tevatron
put more stringent bounds on these couplings as $\mid F_{1R}\mid^{2}<0.30,$
$\mid F_{2L}\mid^{2}<0.05$ and $\mid F_{2R}\mid^{2}<0.12$ assuming
$F_{1L}=1$ at $95\,\%$ C.L. \cite{5}. On the other hand, the indirect
constraint from $b\rightarrow s\gamma$ data by CLEO is $|F_{1R}|<4\times10^{-3}$
at $2\sigma$ level \cite{6}. From $B^{0}-\bar{B}^{0}$ mixings and
rare $B$ decay observables, it is apparent that indirect constraints
are more restrictive than direct constraints for some of the anomalous
couplings \cite{7}.

Extensive studies on the anomalous $Wtb$ couplings described by a
model independent effective Lagrangian approach have been performed
in the literature through single and pair production of top quarks
at hadron colliders \cite{8,9,10,11,12,13,14,15,16,17,18,19,20,21,22}
at lepton colliders \cite{23,24,25,26,27,28,29,30,31,32,33,34} and
at $ep$ colliders \cite{34}.

A high energy electron-proton collider \cite{35} can be realised
by accelerating electrons in a linear accelerator (linac) to 60-140
GeV and colliding them with the 7 TeV protons circulating at the LHC.
When the electron beam is accelerated by a linac, it can be converted
into a beam of high energy real photons, by backscattering off laser
photons. The spectrum of high energy photons would be about 80\% of
the energy of the initial electrons. It has the advantage of obtaining
a 80-90\% polarized electron beam and an intensive high energy photons.
An operation of the LHeC as a $\gamma p$ collider offers interesting
possibilities to study TeV scale physics complementary to its $ep$
option and to the LHC. The production of top quark by FCNC interactions
at the LHeC based $\gamma p$ collider has been studied in \cite{36}.

In this work, we investigate the associated production of single top
quark and $W$ boson through anomalous couplings at the LHeC based
$\gamma p$ collider. It is important to test the couplings of the
top quark with the precision measurement at different energy scale,
which can point the new physics beyond the SM. Therefore, the aim
of this study is to provide bounds on the anomalous couplings of $Wtb$
and $Wtb\gamma$ vertices including corrections from dimension-six
gauge invariant operators at the TeV scale.

\section{Anomalous Interact\i{}ons}

The $Wtb$ coupling is vector-axial (V-A) type in the SM. Therefore,
only the left-handed fermion fields couple to the $W$ boson. The
result of this allows only a left-handed top quark to decay into a
bottom quark and a $W$ boson. However, the new physics can generate
other possible $Wtb$ couplings. The anomalous $Wtb$ couplings can
be expressed in a model independent effective Lagrangian approach
\cite{37}. Furthermore, upon electroweak symmetry breaking there
is another quartic interaction vertex $Wtb\gamma$ giving rise to
the anomalous interactions. We consider the following model independent
effective Lagrangian in the unitary gauge including anomalous $Wtb$
and $Wtb\gamma$ vertices with four independent form factor:

\begin{eqnarray}
L & = & -\frac{g_{W}}{\sqrt{2}}\bar{b}\left[\gamma^{\mu}\left(F_{1L}P_{L}+F_{1R}P_{R}\right)W_{\mu}^{-}\right.\nonumber \\
 &  & \left.+\frac{i\sigma^{\mu\nu}}{2m_{W}}(F_{2L}P_{L}+F_{2R}P_{R})(q_{\nu}W_{\mu}^{-}-q_{\mu}W_{\nu}^{-}+g_{e}(A_{\mu}W_{\nu}^{-}-A_{\nu}W_{\mu}^{-}))\right]t+h.c.\label{eq:1}
\end{eqnarray}
 where left-handed (right-handed) projection operator is $P_{L/R}=\frac{1}{2}(1\mp\gamma_{5})$,
and $\sigma^{\mu\nu}=\frac{i}{2}(\gamma^{\mu}\gamma^{\nu}-\gamma^{\nu}\gamma^{\mu}).$
$q_{\nu}$ is the four-momentum of $W$ boson, $A_{\mu}$($W_{\mu}$)
denotes photon field ($W$ boson field), $F_{iL/R}$ are complex dimensionless
form factors. In the SM, the couplings $F_{1R}=F_{2L}=F_{2R}=0$ and
$F_{1L}$ is equal to CKM matrix element $V_{tb}$, which close to
unity. The same argument can be applied for $F_{1L}=V_{tq}$ where
$q=d,\, s$ when $b$ quark is replaced by $d$ or $s$ quarks in
the Lagrangian. In our calculation, we assume the $F_{iL/R}$'s to
be real for simplicity and define $F_{1L}=\triangle F_{1L}+1$.

For numerical calculations, the anomalous interaction vertices given
in the effective Lagrangian are implemented into the CalcHEP \cite{38}
package. We use the spectrum of photons scattered backward from the
interaction of laser light with the high energy electron beam \cite{39}
and the parton distribution function from CTEQ6M \cite{40} within
this package.

Since top quark decay dominantly via the mode $t\rightarrow Wb$ which
deserves special attention, we first take into account top quark decay
width $\Gamma(t\rightarrow Wb)$ in the presence of anomalous couplings
and find the decay width as:

\begin{eqnarray}
\Gamma(t & \rightarrow & Wb)=\frac{g_{e}^{2}(m_{t}^{2}-m_{W}^{2})^{2}}{64\pi sin^{2}\theta_{W}m_{t}^{3}m_{W}^{2}}\left[(F_{1L}^{2}+2F_{2L}^{2}+F_{1R}^{2}+2F_{2R}^{2})m_{t}^{2}\right.\nonumber \\
 &  & \left.(2F_{1L}^{2}+F_{2L}^{2}+2F_{1R}^{2}+F_{2R}^{2})m_{W}^{2}+6(F_{1R}F_{2L}+F_{1L}F_{2R})m_{t}m_{W}\right]\label{eq:2}
\end{eqnarray}

\begin{figure}
\includegraphics[scale=0.8]{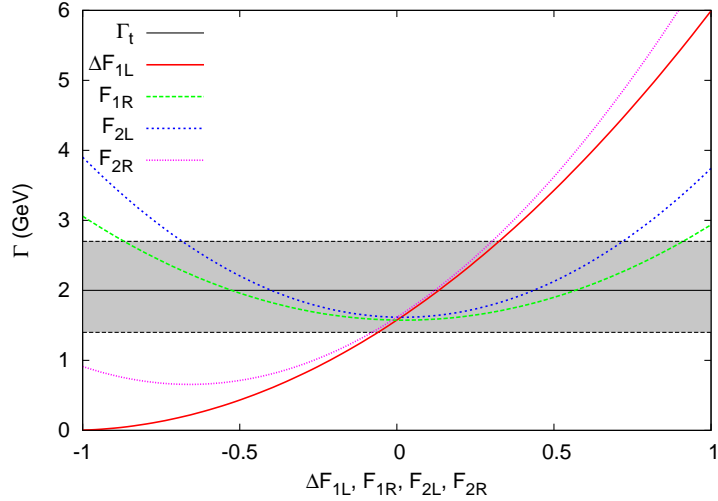}\caption{Decay width of the top quark depending on the anomalous couplings
$F_{1L},\, F_{1R},\, F_{2L}$ and $F_{2R}$.\label{fig:fig1}}
\end{figure}

Fig. \ref{fig:fig1} shows the decay width of top quark depending
on the anomalous couplings by varying one of these couplings at a
time while putting the others equal to zero. The solid line (black)
in this figure denotes the average value of the experimental total
decay width of top quark $\Gamma_{t}=2.0_{-0.6}^{+0.7}$ \cite{41}.
The dashed horizontal lines (black) shows the statistical errors within
$1\sigma$ around the average value. It is seen from Fig. \ref{fig:fig1}
that the limits on the couplings: $\mid F_{1R}\mid<0.5$ and $\mid F_{2L}\mid<0.4$
can be extracted from the intersection point of the experimental average
value of top quark decay width and the theoretical value calculated
with the anomalous couplings. For the couplings $\Delta F_{1L}$ and
$F_{2R}$, assuming positive range, the limits can be found as $\Delta F_{1L}<0.2$
and $F_{2R}<0.2$.

\begin{figure}
\includegraphics{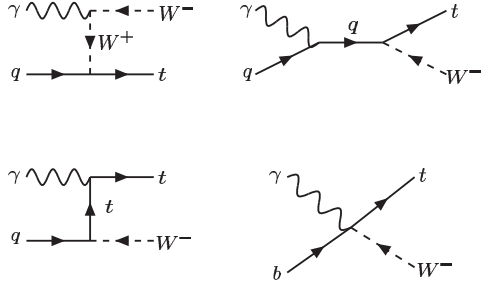}

\caption{ Representative Feynman diagrams for the subprocesses $\gamma q\rightarrow W^{-}t,\,(q=d,s,b$).\label{fig:fig2} }
\end{figure}

The related Feynman diagrams for the subprocess $\gamma q\rightarrow W^{-}t$
are shown in Fig. \ref{fig:fig2}, where $q=d,\, s\,,b$. The last
diagram only contributes to the cross section when initial quark is
a $b$-quark. The differential cross section for the subprocess is
given by the formula

\begin{equation}
\frac{d\hat{\sigma}}{d\hat{t}}=\frac{1}{16\pi\hat{s}^{2}}\sum_{i,j=1,2,3,4}<A_{i}A_{j}^{*}>\label{eq:3}
\end{equation}
 where $<A_{i}A_{j}^{*}>$ is the average over inital state and sum
over final state of the product of amplitudes $A_{i}$ and $A_{j}$
corresponding to the Feynman diagrams given in Fig. \ref{fig:fig2},
and the Mandelstam variables are given as $\hat{s}=(p_{q}+p_{\gamma})^{2}$
and $\hat{t}=(p_{q}-p_{t})^{2}$ in terms of the four-momenta of particles.
The explicit expressions for $A_{i}A_{j}^{*}$ are given in the Appendix.
The total cross section can be obtained by integrating differential
cross section over the parton distribution functions and photon spectrum.

\begin{figure}
\includegraphics[scale=0.8]{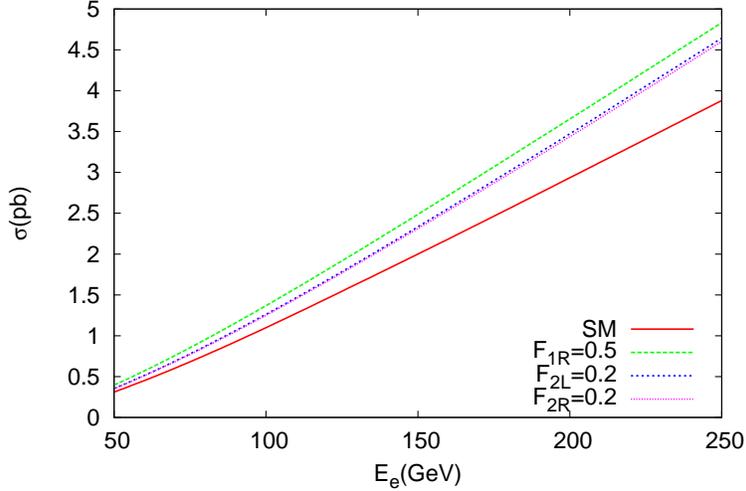}\caption{For the process $\gamma p\rightarrow W^{-}tX$, the dependence of
cross section on the incoming electron beam energy. \label{fig:fig.3}}
\end{figure}

\begin{figure}
\includegraphics[scale=0.8]{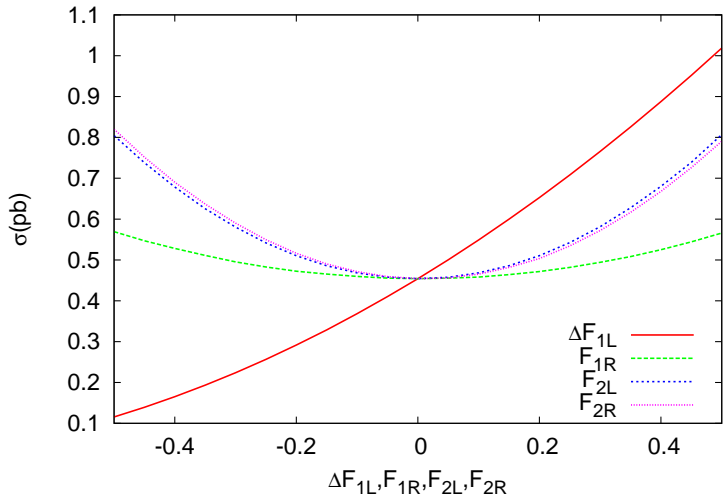}\caption{ For the process $\gamma p\rightarrow W^{-}tX$, the dependence of
cross section on anomalous couplings $\triangle F_{1L},\, F_{1R},\, F_{2L}$
and $F_{2R}$ for electron beam energy of 60 GeV. \label{fig:fig.4} }
\end{figure}

\begin{figure}
\includegraphics[scale=0.8]{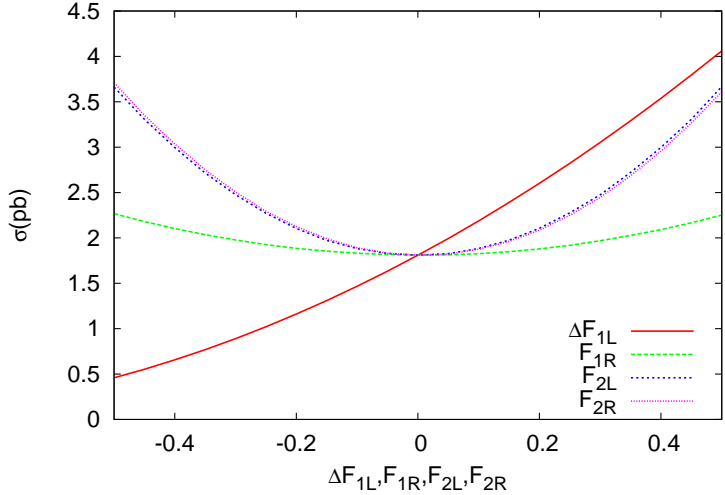}\caption{The same as the Fig. \ref{fig:fig.4} but for electron beam energy
of 140 GeV. \label{fig:fig.5}}
\end{figure}

The total cross section for $\gamma p\rightarrow W^{-}tX$ process
depending on the energy of incoming electron beam is shown in Fig.
\ref{fig:fig.3}. The cross sections for the coupling parameters $F_{1R}=0.5,\, F_{2L}=0.2,\, F_{2R}=0.2$
are larger than the cross section for the SM. In plotting Figs. \ref{fig:fig.3},
\ref{fig:fig.4} and \ref{fig:fig.5}, one anomalous parameter is
kept nonzero while the others are zero.

In Figs. \ref{fig:fig.4} and \ref{fig:fig.5}, we present total cross
section of $\gamma p\to W^{-}tX$ as a function of anomalous couplings
$\Delta F_{1L}$, $F_{1R}$, $F_{2L}$ and $F_{2R}$ with taking the
energy of incoming electron to be $E_{e}$=60 GeV and $E_{e}$=140
GeV, respectively.

We calculate the cross section corresponding to the SM case for the
anomalous couplings $\Delta F_{1L}=0$, $F_{1R}=0,$ $F_{2L}=0$ and
$F_{2R}=0$. From Figs. \ref{fig:fig.4} and \ref{fig:fig.5}, it
is seen that the cross sections have minimum when $F_{1R}=0,$ $F_{2L}=0$
and $F_{2R}=0$, corresponding the values $0.47$ pb and $1.809$
pb for the center of mass energy $\sqrt{s_{ep}}=1.29$ TeV and $\sqrt{s_{ep}}=1.98$
TeV, respectively. With these figures, we observe the cross section,
which shows symmetric behaviour around zero for the couplings $F_{1R}$,
$F_{2L}$ and $F_{2R}$ has approximately the same dependence to $F_{2L}$
and $F_{2R}$ couplings, and also the cross section enhances with
increasing $\Delta F_{1L}$. The difference between the cross sections,
when we set $F_{2L}=0.5$ or $F_{2R}=0.5$ and the SM value is about
$75\%$ as shown in Fig. \ref{fig:fig.4}, while it is $100\%$ as
in Fig. \ref{fig:fig.5}. However, the cross sections changes slightly
with $F_{1R}$.

\section{sens\i{}t\i{}v\i{}ty to anomalous coupl\i{}ngs}

In this section, the sensitivity to the anomalous $Wtb$ couplings
is discussed taking into account the subprocesses $\gamma q\rightarrow W^{-}t$
and $\gamma\bar{q}\rightarrow W^{+}\bar{t}$, where $q=d,s,b$. We
estimate the sensitivity to these anomalous couplings at LHeC based
$\gamma p$ colliders for the integrated luminosities of 1, 10 and
100 $fb^{-1}.$ We use $\chi^{2}$ function to obtain sensitivity:

\begin{equation}
\chi^{2}=\left(\frac{\sigma_{SM}-\sigma(\Delta F_{1L},F_{1R},F_{2L},F_{2R})}{\Delta\sigma_{SM}}\right)^{2}\label{eq:eq.4}
\end{equation}
 where $\Delta\sigma_{SM}=\sigma_{SM}\sqrt{\delta_{stat.}^{2}}$ with
$\delta_{stat.}=1/\sqrt{N_{SM}}$, $N_{SM}$ is the number of events
calculated by $N_{SM}=\sigma_{SM}\times BR(t\rightarrow W^{+}b)\times BR(W^{+}\rightarrow hadrons)\times BR(W^{-}\rightarrow l^{-}\nu)\times\epsilon_{b-tag}\times L_{int}$.
We take into account $W^{+}$ boson decays hadronically while $W^{-}$
decays leptonically corresponding to relevant branchings and assume
b-jet tagging efficiency as $50\%$. In our calculations, we consider
only one of the couplings is assumed to deviate from its SM value
at a time.

The limits for the anomalous coupling parameters of top quark are
given in Table \ref{tab:tab1} for integrated luminosities $L_{int}=1,\,10,\,100$
fb$^{-1}$ and electron beam energy of 60 GeV. Table \ref{tab:tab2}
presents the limits on these couplings for electron beam energy of
140 GeV. In these tables, the limits for the integrated luminosities
of $10$ fb$^{-1}$ and $100$ fb$^{-1}$ are much better than the
LHC results at $\sqrt{s}=7$ TeV \cite{1,3} and Tevatron results
\cite{5}. The expected precision limits of the anomalous top quark
coupling measurements has been estimated as $\mid\Delta F_{1L}\mid\lesssim0.1$,
$\mid F_{1R}\mid\lesssim0.3$$,\,\mid F_{2L}\mid\lesssim0.15\mbox{ and \ensuremath{\mid F_{2R}\mid\lesssim0.024}}$
\cite{42} at the LHC with $\sqrt{s}=14$ TeV and $L_{int}=10\, fb^{-1}$.
The limits on the anomalous couplings found in this study can also
be compared with the limits obtained from photon induced process in
hadron-hadron collisions at the LHC \cite{43}. The studies on anomalous
$Wtb$ couplings have shown the sensitivity for the limits of $-0.02\lesssim F_{2L}\lesssim0.06$
and $\mid F_{2R}\mid\lesssim0.05$ at linear collider with $\sqrt{s}=500$
GeV and $L_{int}=500\, fb^{-1}$ \cite{44,45}. In this study, we
find much better limits on the anomalous couplings of top quark as
-0.0187<$\Delta F_{1L}$< 0.0172, -0.1884<$F_{1R}$<0.1957, -0.1014<$F_{2L}$<0.0939
and -0.0871<$F_{2R}$<0.1058 at an integrated luminosity of $10$
fb$^{-1}$ for the electron beam energy of 140 GeV at the LHeC based
$\gamma p$ collider.

\begin{table}
\caption{Sensitivity (95\% C.L.) to anomalous $Wtb$ couplings at the LHeC
based $\gamma p$ collider with electron beam energy of 60 GeV for
various integrated luminosities. \label{tab:tab1}}

\begin{tabular}{|c|c|c|c|c|}
\hline
$L(fb^{-1})$  & $\Delta F_{1L}$  & $F_{1R}$  & $F_{2L}$  & $F_{2R}$\tabularnewline
\hline
\hline
1  & -0.1088: +0.1318  & -0.5258: +0.5328  & -0.3010: +0.2995  & -0.2903: +0.3106\tabularnewline
\hline
10  & -0.0187: +0.055  & -0.3350: +0.3422  & -0.1923: +0.1901  & -0.1802: +0.2035\tabularnewline
\hline
100  & -0.0065: +0.0314  & --0.1082: +0.1188  & -0.0601: +0.0626  & -0.1233: +0.1579\tabularnewline
\hline
\end{tabular}
\end{table}

\begin{table}
\caption{The same as the Table \ref{tab:tab1}, but for electron beam energy
of 140 GeV.\label{tab:tab2} }

\begin{tabular}{|c|c|c|c|c|}
\hline $L(fb^{-1})$  & $\Delta F_{1L}$  & $F_{1R}$  & $F_{2L}$  &
$F_{2R}$\tabularnewline \hline \hline 1  & -0.0617: +0.0581 &
-0.3416: +0.3490  & -0.1705: +0.1684  & -0.1643:
+0.1781\tabularnewline \hline 10  & -0.0187: +0.0172  & -0.1884:
+0.1957  & -0.1014: +0.0939  & -0.0871 : +0.1058\tabularnewline
\hline 100  & -0.0022: +0.0074  & -0.1020: + 0.1118  & -0.0524:
+0.0554  & -0.0453 : +0.0652\tabularnewline \hline
\end{tabular}
\end{table}

\section{Conclus\i{}ons}

The precision measurement of the anomalous couplings relevant to $Wtb$
and $Wtb\gamma$ vertices is very important for the future experiments,
and it can point the new physics beyond the SM. The limits on the
couplings $\Delta F_{1L}$ and $\mbox{\ensuremath{F_{1R}}}$ are better
than the limits from direct searches at Tevatron and LHC. However,
the limits $F_{2L}$ and $\mbox{\ensuremath{F_{2R}}}$ become comparable
to the results from LHC and high energy linear colliders. The measurement
of single top production at LHeC based $\gamma p$ collider would
provide complementary information to the LHC data that could help
in determining anomalous $Wtb$ couplings. In addition to QCD explorer
searches, the LHeC will contribute to top quark physics at high momentum
transfer in deep inelastic scattering.
\begin{acknowledgments}
Authors would like to thank O. Cakir for valuable suggestions and
comments.
\end{acknowledgments}

\section*{Appendix}

The explicit expressions for $<A_{i}A_{j}^{*}>$ corresponding to
Feynman diagrams in Fig. \ref{fig:fig2} are following:

\begin{eqnarray*}
<\left|A_{1}\right|^{2}> & = & -\frac{1}{16m_{W}^{5}sin\theta_{W}^{2}(m_{W}^{2}-t)^{2}}g{}_{e}^{4}((F_{2L}F_{1R}+F_{1L}F_{2R})m_{t}(m{}_{W}^{6}(20s-13t)-t^{3}(2s+t)\\
 &  & -m_{W}^{2}t^{2}(4s+13t)-m_{W}^{4}t(22s+47t)+m_{t}^{2}(13m_{W}^{6}+47m_{W}^{4}t-13m_{W}^{2}t^{2}+t^{3}))\\
 &  & +(F_{2L}^{2}+F_{2R}^{2})m_{W}(4m_{t}^{4}(m_{W}^{4}+4m_{W}^{2}t-t^{2})+t(5(m_{W}^{3}-2m_{W}s)^{2}\\
 &  & +(-13m_{W}^{4}+24m_{W}^{2}s-4s^{2})t-(m_{W}^{2}+4s)t^{2}+t^{3})\\
 &  & +m_{t}^{2}(-5m_{W}^{6}+t^{2}(4s+3t)-3m_{W}^{2}t(8s+5t)+m_{W}^{4}(20s+9t)))\\
 &  & +2(F_{1L}^{2}+F_{1R}^{2})m_{W}(m_{t}^{4}(7m_{W}^{4}-4m_{W}^{2}t+t^{2})+m_{t}^{2}(2m_{W}^{6}+2m_{W}^{2}t^{2}+m_{W}^{4}(s+t)\\
 &  & -t^{2}(s+t))+2m_{W}^{2}(m_{W}^{4}(5s-t)+t(s^{2}+st+t^{2})-m_{W}^{2}(5s^{2}+6st+4t^{2}))))
\end{eqnarray*}

\begin{eqnarray*}
2Re<A_{1}A_{2}^{*}> & = & (g{}_{e}^{4}(m_{W}^{2}(F_{2R}^{2}m_{t}^{2}(8m_{t}^{4}-5m_{W}^{4}+2m_{t}^{2}(7m_{W}^{2}-6s)+5m_{W}^{2}s+4s^{2})\\
 &  & +F_{1R}^{2}(4m_{t}^{6}+m_{t}^{4}(19m_{W}^{2}-7s)-m_{t}^{2}s(7m_{W}^{2}+s)+2s(5m_{W}^{4}-6m_{W}^{2}s+s^{2})))\\
 &  & -(F_{2R}^{2}m_{W}^{2}(10m_{t}^{4}-5m_{W}^{4}+3m_{t}^{2}s+11m_{W}^{2}s-4s^{2})\\
 &  & +F_{1R}^{2}(m_{t}^{4}(12m_{W}^{2}-s)+4m_{W}^{2}(5m_{W}^{2}-s)s+m_{t}^{2}(7m_{W}^{4}-7m_{W}^{2}s+s^{2})))t\\
 &  & +(F_{2R}^{2}m_{W}^{2}(-3m_{t}^{2}-14m_{W}^{2}+9s)+F_{1R}^{2}(m_{t}^{4}+m_{t}^{2}(4m_{W}^{2}-2s)+6m_{W}^{2}(-2m_{W}^{2}+s)))t^{2}\\
 &  & +(5F_{2R}^{2}m_{W}^{2}-F_{1R}^{2}(m_{t}^{2}-4m_{W}^{2}))t^{3}+(F_{2L}F_{1R}+F_{1L}F_{2R})m_{t}m_{W}(m_{W}^{4}(3s-20t)\\
 &  & +2m_{t}^{4}(11m_{W}^{2}+t)-3(s-2t)t(s+t)+m_{W}^{2}(s^{2}-6st-10t^{2})\\
 &  & +4m_{t}^{2}(5m_{W}^{4}-2t^{2}-3m_{W}^{2}(2s+t)))+F_{2L}^{2}m_{W}^{2}(8m_{t}^{6}+2m_{t}^{4}(7m_{W}^{2}-6s-5t)\\
 &  & +m_{t}^{2}(-5m_{W}^{4}+5m_{W}^{2}s+4s^{2}-3st-3t^{2})+t(5m_{W}^{4}+(s+t)(4s+5t)-m_{W}^{2}(11s+14t)))\\
 &  & +F_{1L}^{2}(4m_{t}^{6}m_{W}^{2}+m_{t}^{4}(19m_{W}^{4}+t(s+t)-m_{W}^{2}(7s+12t))-\\
 &  & -m_{t}^{2}(7m_{W}^{4}(s+t)+t(s+t)^{2}+m_{W}^{2}(s^{2}-7st-4t^{2}))+2m_{W}^{2}(5m_{W}^{4}s\\
 &  & +(s+t)(s^{2}+st+2t^{2})-2m_{W}^{2}(3s^{2}+5st+3t^{2})))))/(12m_{W}^{4}sin^{2}\theta_{W}(m_{W}^{2}-t)(m_{W}^{2}-s-t))
\end{eqnarray*}

\begin{eqnarray*}
2Re<A_{1}A_{3}^{*}> & = & \frac{1}{24m_{W}^{4}sin\theta_{W}^{2}s(m_{W}^{2}-t)}g{}_{e}^{4}(m{}_{W}^{2}(F_{2R}^{2}m_{t}^{2}(-8m_{t}^{4}+6m_{t}^{2}(m_{W}^{2}+2s)+(2m_{W}^{2}-s)(m_{W}^{2}+4s))\\
 &  & +F_{1R}^{2}(-4m_{t}^{6}+2(m_{W}^{2}-s)s(4m_{W}^{2}+s)+m_{t}^{4}(-4m_{W}^{2}+7s)+m_{t}^{2}(8m_{W}^{4}+4m_{W}^{2}s+s^{2})))\\
 &  & +F_{2R}^{2}m_{W}^{2}(6m_{t}^{4}-2m_{W}^{4}+3m_{W}^{2}s+4s^{2}-m_{t}^{2}(4m_{W}^{2}+9s))\\
 &  & +F_{1R}^{2}(m_{t}^{4}+(4m_{W}^{2}-s)-2m_{W}^{2}(4m_{W}^{2}+s^{2})+m_{t}^{2}(4m_{W}^{4}-3m_{W}^{2}s+s)))t\\
 &  & +F_{1R}^{2}(m_{t}^{2}-4m_{W}^{2})s-F_{2R}^{2}m_{W}^{2}(-2m_{t}^{2}+2m_{W}^{2}+s))t^{2}+(F_{2L}F_{1R}+F_{1L}F_{2R})m_{t}m_{W}\\
 &  & (2m_{W}^{4}(8s-11t)+s(3s-2t)t-2m_{t}^{4}(11m_{W}^{2}+t)-m_{W}^{2}(s^{2}+14st+2t^{2})\\
 &  & +2m_{t}^{2}(11m_{W}^{4}+t^{2}+12m_{W}^{2}(s+t)))+F_{2L}^{2}m_{W}^{2}(-8m_{t}^{6}+6m_{t}^{4}(m_{W}^{2}+2s+t)\\
 &  & +m_{t}^{2}(2m_{W}^{4}-4s^{2}+m_{W}^{2}(7s-4t)-9st+2t^{2})-t(2m_{W}^{4}+s(-4s+t)+m_{W}^{2}(-3s+2t)))\\
 &  & +F_{1L}^{2}(-4m_{t}^{6}m_{W}^{2}+m_{t}^{2}(8m_{W}^{6}+m_{W}^{2}s(s-3t)+4m_{W}^{4}(s+t)+st(s+t))\\
 &  & +m_{t}^{4}(-4m_{W}^{4}-st+m_{W}^{2}(7s+4t))+(2m_{W}^{2}(-3m_{W}^{2}s^{2}+4m_{W}^{4}(s-t)-s(s^{2}+st+2t^{2}))))
\end{eqnarray*}

\begin{eqnarray*}
2Re<A_{1}A_{4}^{*}> & = & \frac{1}{16m_{W}^{5}sin\theta_{W}^{2}(m_{W}^{2}-t)}g{}_{e}^{4}((F_{2L}F_{1R}+F_{2R}F_{1L})mt\\
 &  & (-12m_{W}^{2}t(s+t)+t^{2}(2s+t)+m_{W}^{4}(2s+35t)-m{}_{t}^{2}(35m_{W}^{4}-12m_{W}^{2}t+t^{2}))\\
 &  & +(F_{2L}^{2}+F_{2R}^{2})m_{W}(8m_{t}^{4}(-3m_{W}^{2}+t)+t((m_{W}^{2}-2s)^{2}+4(2m_{W}^{2}+s)t-t^{2})\\
 &  & -m{}_{t}^{2}(m_{W}^{4}-4m_{W}^{2}(s+4t)+t(12s+7t))))
\end{eqnarray*}

\begin{eqnarray*}
\left|A_{2}\right|^{2} & =- & \frac{1}{(9m_{W}^{2}sin\theta_{W}^{2}(-m_{W}^{2}+s+t)^{2})}g{}_{e}^{4}(2(F_{1R}^{2}+2F{}_{2R}^{2})m{}_{t}^{6}+6(F_{1R}^{2}+F_{2R}^{2})m{}_{t}^{4}m{}_{W}^{4}\\
 &  & -(2F_{1R}^{2}+F_{2R}^{2})m_{t}^{2}m_{W}^{4}-4(F_{1R}^{2}+2F_{2R}^{2})m_{t}^{4}s-(5F_{1R}^{2}+4F_{2R}^{2})m_{t}^{2}m_{W}^{2}s\\
 &  & +2(F_{1R}^{2}+2F_{2R}^{2})m_{W}^{4}s+2(F_{1R}^{2}+2F{}_{2R}^{2})m_{t}^{2}s^{2}-(2F_{1R}^{2}+F{}_{R2}^{2})m_{W}^{2}s^{2}+\\
 &  & +6(F_{2L}F_{1R}+F_{1L}F_{2R})m_{t}m_{W}(m_{t}^{2}-s-t)(2m_{t}^{2}+m_{W}^{2}-s-t)\\
 &  & +(F_{2R}^{2}(-8m_{t}^{4}+2m_{W}^{4}-2m_{t}^{2}(m_{W}^{2}-3s)-5m_{W}^{2}s+2s^{2})+F_{1R}^{2}\\
 &  & (-4m_{t}^{4}+m_{W}^{4}-4m_{W}^{2}s+s^{2}+m_{t}^{2}(-4m_{W}^{2}+3s)))t\\
 &  & +(F_{1R}^{2}+2F_{2R}^{2})(m_{t}^{2}-2m_{W}^{2}+2s+t)t^{2}+F_{2L}^{2}(4m_{t}^{6}+m_{t}^{4}(6m_{W}^{2}-8(s+t))\\
 &  & -2m_{t}^{2}(m_{W}^{4}+m_{W}^{2}(2s+t)-(s+t)(2s+t))+(m_{W}^{2}-s-t)(-2t(s+t)+m_{W}^{2}(s+2t)))\\
 &  & +F_{1L}^{2}(2m_{t}^{6}+m_{t}^{4}(6m_{W}^{2}-4(s+t))+(m_{W}^{2}-s-t)(-t(s+t)+m_{W}^{2}(2s+t))\\
 &  & +m_{t}^{2}(-m_{W}^{4}+(s+t)(2s+t)-m_{W}^{2}(5s+4t))))
\end{eqnarray*}

\begin{eqnarray*}
2Re<A_{2}A_{3}^{*}> & = & -\frac{1}{9m_{W}^{2}sin\theta_{W}^{2}s(-m_{W}^{2}+s+t)}g{}_{e}^{4}(F_{1R}^{2}m_{t}^{4}(m_{t}^{2}+m_{W}^{2})+F_{2R}M{}_{t}^{4}(2m_{t}^{2}-m_{W}^{2})\\
 &  & -2F_{1R}^{2}m_{t}^{2}(m_{W}^{4}+m_{t}^{2}s)-F_{2R}^{2}m_{t}^{2}(m_{W}^{4}+4m_{t}^{2}s)-(F_{1R}^{2}+F_{2R}^{2})m_{t}^{2}m_{W}^{2}s+F_{2R}^{2}m_{W}^{2}s(s-m_{W}^{2})\\
 &  & +(F_{1R}^{2}+2F_{2R}^{2})m_{t}^{2}s^{2}+(-F_{2R}^{2}(2m_{t}^{2}+m_{W}^{2})(m_{t}^{2}-m_{W}^{2}-s)-F_{1R}^{2}(m_{t}^{2}+m_{t}^{4}m_{W}^{2}\\
 &  & -2m_{W}^{4}+m_{W}^{2}s-s^{2}))t+F_{1R}^{2}st^{2}+(F_{2L}F_{1R}+F_{1L}F_{2R})m_{t}m_{W}(6m_{t}^{4}-m_{W}^{2}(s-6t)\\
 &  & -3m_{t}^{2}(2m_{W}^{2}+3s+2t)+s(3s+4t))+F_{1L}^{2}(m_{t}^{6}+m_{t}^{4}(m_{W}^{2}-2s-t)\\
 &  & -m_{t}^{2}(2m_{W}^{4}-s^{2}+m_{W}^{2}(s+t))+t(2m_{W}^{4}-m_{W}^{2}s+s(s+t)))\\
 &  & +F_{2L}^{2}(2m_{t}^{6}-m_{t}^{2}(m_{W}^{2}+2s)(m_{W}^{2}-s-t)-m_{t}^{4}(m_{W}^{2}+4s+2t)+m_{W}^{2}(m_{W}^{2}(-s+t)+s(s+t))))\\
\end{eqnarray*}

\begin{eqnarray*}
2Re<A_{2}A_{4}^{*}> & = & \frac{1}{6m_{W}^{3}sin\theta_{W}^{2}(m_{W}^{2}-s-t)}g{}_{e}^{4}((F_{2L}^{2}+F_{2R}^{2})m_{W}(5m_{t}^{4}-2(s+t)^{2}+m_{W}^{2}(2s+t)\\
 &  & -m_{t}^{2}(m_{W}^{2}+3(s+t)))+(F_{2L}F_{1R}+F_{2R}F_{1L})m_{t}(m_{t}^{4}+(s+t)(s+3t)\\
 &  & -m_{W}^{2}(7s+8t)+m_{t}^{2}(8m_{W}^{2}-2(s+2t))))
\end{eqnarray*}

\begin{eqnarray*}
\left|A_{3}\right|^{2} & = & \frac{1}{36m_{W}^{2}sin\theta_{W}^{2}s}g{}_{e}^{4}(m_{W}^{2}(-(F_{1L}^{2}+2F_{2L}^{2}+F_{1R}^{2}+2F_{2R}^{2})m_{t}^{2}-6(F_{2L}F_{1R}+F_{1L}F_{2R})m_{t}m_{W}\\
 &  & +(2F_{1L}^{2}+F_{2L}^{2}+2F_{1R}^{2}+F_{2R}^{2})(s-m_{W}^{2}))+((F_{1R}^{2}+2F_{2R}^{2})m_{t}^{2}+6(F_{2L}F_{1R}+F_{1L}F_{2R})m_{t}m_{W}\\
 &  & +(2F_{1R}^{2}+F_{2R}^{2})m_{W}^{2}+F_{2L}^{2}(2m_{t}^{2}+m_{W}^{2}-2s)+F_{1L}^{2}(m_{t}^{2}+2m_{W}^{2}-s)\\
 &  & -(F_{1R}^{2}+2F_{2R}^{2})s)t)
\end{eqnarray*}

\begin{eqnarray*}
2Re<A_{3}A_{4}^{*}> & =- & \frac{1}{12m_{W}^{3}s^{2}sin\theta_{W}^{2}}g{}_{e}^{4}((F_{2L}F_{1R}+F_{2R}F_{1L})m_{t}(m_{t}^{4}+m_{W}^{2}(-4s+t)+s(s+t)\\
 &  & -m_{t}^{2}(m_{W}^{2}+2s+t))+(F_{2L}^{2}+F_{2R}^{2})m_{W}(m_{t}^{4}+2s^{2}+m_{W}^{2}(-2s+t)\\
 &  & -m_{t}^{2}(m_{W}^{2}+3s+t)))
\end{eqnarray*}

\[
\left|A_{4}\right|^{2}=\frac{1}{8m_{W}^{4}sin\theta_{W}^{2}}g{}_{e}^{4}(F_{2L}^{2}+F_{2R}^{2})(2m_{t}^{4}+2s(s+t)-m_{W}^{2}(2s+t)+m_{t}^{2}(m_{W}^{2}-2(2s+t)))
\]


\begin{thebibliography}{10}
\bibitem[1]{1} CMS Collaboration, note CMS-PAS-TOP-10-008. G. Aad
\emph{et al.} {[}Atlas Collaboration{]}, note ATLAS-CONF- 2011-027.

\bibitem[2]{2} G. Aad \emph{et al}. {[}Atlas Collaboration{]}, note
ATLAS-CONF- 2011-027.

\bibitem[3]{3} G. Aad \emph{et al}. {[}Atlas Collaboration{]}, note
ATLAS-CONF- 2011-037.

\bibitem[4]{4} J. A. Aguilar-Saavedra, N. F. Castro and A. Onofre,
Phys. Rev. D \textbf{83}, 117301 (2011) {[}arXiv:1105.0117 {[}hep-ph{]}{]}.

\bibitem[5]{5} V. M. Abazov \emph{et al}. {[}D0 Collaboration{]},
Phys. Lett. B \textbf{713}, 165 (2012) {[}arXiv:1204.2332 {[}hep-ex{]}{]}.

\bibitem[6]{6} F. Larios, M. A. Perez and C. P. Yuan, Phys. Lett.
B \textbf{457}, 334 (1999) {[}hep-ph/9903394{]}.

\bibitem[7]{7} J. Drobnak, S. Fajfer, and J. F. Kamenik, Nucl. Phy.\textbf{
}B \textbf{855}, 82 (2012).

\bibitem[8]{8} D.~O.~Carlson and C.~P.~Yuan, Phys.\ Lett.\ B
\textbf{306}, 386 (1993).

\bibitem[9]{9} D.~O.~Carlson, E.~Malkawi and C.~P.~Yuan, Phys.\ Lett.\ B
\textbf{337}, 145 (1994) {[}hep-ph/9405277{]}.

\bibitem[10]{10} T.~Stelzer and S.~Willenbrock, Phys.\ Lett.\ B
\textbf{357}, 125 (1995) {[}hep-ph/9505433{]}.

\bibitem[11]{11} A.~Heinson, A.~S.~Belyaev and E.~E.~Boos, Phys.\ Rev.\ D
\textbf{56}, 3114 (1997) {[}hep-ph/9612424{]}.

\bibitem[12]{12} E.~Boos, L.~Dudko and T.~Ohl, Eur.\ Phys.\ J.\ C
\textbf{11}, 473 (1999) {[}hep-ph/9903215{]}.

\bibitem[13]{13} F.~del Aguila and J.~A.~Aguilar-Saavedra, Phys.\ Rev.\ D
\textbf{67}, 014009 (2003) {[}hep-ph/0208171{]}.

\bibitem[14]{14} M.~M.~Najafabadi, JHEP \textbf{0803}, 024 (2008)
{[}arXiv:0801.1939 {[}hep-ph{]}{]}.

\bibitem[15]{15} J.~A.~Aguilar-Saavedra, Nucl.\ Phys.\ B \textbf{804},
160 (2008) {[}arXiv:0803.3810 {[}hep-ph{]}{]}.

\bibitem[16]{16} S.~K.~Gupta, A.~S.~Mete and G.~Valencia, Phys.\ Rev.\ D
\textbf{80}, 034013 (2009) {[}arXiv:0905.1074 {[}hep-ph{]}{]}.

\bibitem[17]{17} E.~L.~Berger, Q.~-H.~Cao and I.~Low, 
 Phys.\ Rev.\ D \textbf{80}, 074020 (2009) {[}arXiv:0907.2191 {[}hep-ph{]}{]}.

\bibitem[18]{18} C.~Zhang and S.~Willenbrock, 
 Phys.\ Rev.\ D \textbf{83}, 034006 (2011) {[}arXiv:1008.3869 {[}hep-ph{]}{]}.

\bibitem[19]{19} S.~D.~Rindani and P.~Sharma, 
 JHEP \textbf{1111}, 082 (2011) {[}arXiv:1107.2597 {[}hep-ph{]}{]}.

\bibitem[20]{20} S.~D.~Rindani and P.~Sharma, 
 Phys.\ Lett.\ B \textbf{712}, 413 (2012) {[}arXiv:1108.4165 {[}hep-ph{]}{]}.

\bibitem[21]{21} K.~Kolodziej, 
 Phys.\ Lett.\ B \textbf{710}, 671 (2012) {[}arXiv:1110.2103 {[}hep-ph{]}{]}.

\bibitem[22]{22} K.~Kolodziej, 
 arXiv:1212.6733 {[}hep-ph{]}. 


\bibitem[23]{23} S.~Ambrosanio and B.~Mele, 
 Z.\ Phys.\ C \textbf{63}, 63 (1994) {[}hep-ph/9311263{]}.

\bibitem[24]{24} N.~V.~Dokholian and G.~V.~Jikia, 
 Phys.\ Lett.\ B \textbf{336}, 251 (1994).

\bibitem[25]{25} K.~Hagiwara, M.~Tanaka and T.~Stelzer, 
 Phys.\ Lett.\ B \textbf{325}, 521 (1994) {[}hep-ph/9401295{]}.

\bibitem[26]{26} E.~Boos, M.~Sachwitz, H.~J.~Schreiber, S.~Shichanin,
A.~Pukhov, V.~Ilin, T.~Ishikawa and T.~Kaneko \textit{et al.},
 Phys.\ Lett.\ B \textbf{326}, 190 (1994).

\bibitem[27]{27} E.~Boos, M.~Dubinin, M.~Sachwitz and H.~J.~Schreiber,
 Eur.\ Phys.\ J.\ C \textbf{16}, 269 (2000) {[}hep-ph/0001048{]}.

\bibitem[28]{28} B.~Grzadkowski and Z.~Hioki, 
 Nucl.\ Phys.\ B \textbf{585}, 3 (2000) {[}hep-ph/0004223{]}.

\bibitem[29]{29} E.~Boos, M.~Dubinin, A.~Pukhov, M.~Sachwitz
and H.~J.~Schreiber, 
 Eur.\ Phys.\ J.\ C \textbf{21}, 81 (2001) {[}hep-ph/0104279{]}.

\bibitem[30]{30} K.~Cieckiewicz and K.~Kolodziej, 
 Acta Phys.\ Polon.\ B \textbf{34}, 5497 (2003) {[}hep-ph/0310300{]}.

\bibitem[31]{31} K.~Kolodziej, 
 Phys.\ Lett.\ B \textbf{584}, 89 (2004) {[}hep-ph/0312168{]}.

\bibitem[32]{32} P.~Batra and T.~M.~P.~Tait, 
 Phys.\ Rev.\ D \textbf{74}, 054021 (2006) {[}hep-ph/0606068{]}.

\bibitem[33]{33} B.~Sahin and I.~Sahin, 
 Eur.\ Phys.\ J.\ C \textbf{54}, 435 (2008) {[}arXiv:0709.0365
{[}hep-ph{]}{]}.

\bibitem[34]{34} S.~Atag, O.~Cakir and B.~Dilec, 
 Phys.\ Lett.\ B \textbf{522}, 76 (2001) {[}hep-ph/0107179{]}.

\bibitem[35]{35} J. L. Abelleira Fernandez \emph{et al}., LHeC Study
Group, J. Phy. G: Nucl. Part. Phys. \textbf{39}, 075001 (2012).

\bibitem[36]{36} \.{I}. T. Cakir, O. Cakir and S. Sultansoy, \emph{et
al}. Phys. Let. B \textbf{685}, 170 (2010).

\bibitem[37]{37} J. A. Aguilar-Saavedra, Nucl. Phys. B \textbf{812},
181 (2009); J. M. Yang, B. Young, Phys. Rev. D \textbf{56}, 5907 (1997).

\bibitem[38]{38} A. Pukhov\emph{ et al.}, arXiv:hep-ph/9908288, (1999);
A. Pukhov, arXiv:hep-ph/0412191, (2004).

\bibitem[39]{39} I. F.Ginzburg, G.L.Kotkin, V.G.Serbo and V.I.Telnov,
Nucl. Instrum. Meth. \textbf{205}, 47 (1983).

\bibitem[40]{40} J. Pumplin\emph{ et al}., JHEP \textbf{0207, }012
(2002). arXiv:hep-ph/0201195.

\bibitem[41]{41} J. Beringer \emph{et al.} {[}Particle Data Group
Collaboration{]}, Phys. Rev. D \textbf{86}, 010001 (2012).

\bibitem[42]{42} F.Bach, T. Ohl, Phys. Rev. D \textbf{86,} 114026
(2012).

\bibitem[43]{43} B. Sahin, A. Billur, Phys.Rev. D \textbf{86} 074026
(2012).

\bibitem[44]{44} E. Devetak, A. Nomerotski, Phys. Rev. D \textbf{84},
034029 (2011).

\bibitem[45]{45} E. Boos \emph{et al}., Eur. Phys. J. C \textbf{21},
81\textendash{}91 (2001). \end{thebibliography}
\end{document}